\documentclass[showpacs,preprintnumbers,amsmath,amssymb]{revtex4}
\usepackage{graphicx}
\usepackage{bm}

\begin{document}

\title{Scaling of entanglement in finite arrays of exchange-coupled quantum dots}

\author{Jiaxiang Wang and Sabre Kais}
\affiliation{Chemistry Department, Purdue University, West 
Lafayette, IN 47907}

\begin{abstract}
We present a  finite-size scaling analysis of 
the entanglement in a two-dimensional arrays of quantum dots 
modeled by the Hubbard Hamiltonian on a
triangular lattice. Using multistage block renormalization group approach,
we have  found that there is an abrupt jump of the entanglement
when  a first-order quantum phase transition occurs.
At the critical point, the entanglement is constant, independent of the block size. 
\end{abstract}
\pacs{03.65.Ud,71.10.Fd,05.10.Cc}

\maketitle

\section{Introduction}
After the pioneering work of Benioff \cite{benioff},
 Feynman\cite{feynman}, 
and Deutsch \cite{deutsch}  in field of
quantum information  and  computing, the development in this field has been
explosive in the past few years \cite{bennett, monroe}. 
A lot of exciting progresses have been
made, such as the experimental realization of Shor's quantum factoring
algorithm with NMR scheme for quantum computation \cite{vandersypen}, 
teleportation \cite{bouwmeester},
cryptography \cite{jennewein} and dense coding \cite{Mattle}. 
Central to all these remarkable achievements
is the concept of the quantum entanglements 
\cite{Lewenstein,Horodecki,donald,barbara,bruss}. Just like energy, the entanglement has now been regarded as
a controllable and a fungible physical resource \cite{nielsen}. Almost all the efficient
protocols in quantum computing applications are based upon its generation.
Experimentally,  entanglements have already been produced among up to four
photons \cite{lamas,eibl} and even between two macroscopic states such as two superconducting
qubits, each of which contains as large as $10^{9}$ electrons \cite{berkley}. But
theoretically, because an ensemble's Hilbert space grows exponentially with
the number of its component particles, we are still far from fully
understanding the contents of the entanglements. Only for the simplest state
with two distinguishable particles can we have a complete description of the
entanglement measure. For states of more than two particles, especially for
mixed states, the current knowledge about their entanglement is very
limited and all the related complexities have just begun to be explored.

\section{Entanglement measure for fermions}

For spin-only entanglement of localized distinguishable particles, the most 
suitable and famous measure of the entanglement is the Wootters' measure \cite{wootters}. 
Recently, Schlieman\cite{schliemann,schliemann1} examined the 
influence of quantum statistics upon the definition of the entanglement.
He discussed a two-fermion system with the Slater decomposition instead of
Schmit decomposition for the entanglement measure. If we take  each of
the indistinguishable fermions to be in the single-particle Hilbert space $C^N$ with
$f_m,f_m^{+} (m=1,...,N)$ denotes the fermionic annihilation and creation
operator of single-particle states and 
$\left| \Omega \right\rangle $ represents the vacuum state. Then a pure
two-electron state can be written as $\sum\limits_{m,n}^{} {\omega _{mn} } f_m ^ +  f_n ^ +  \left| \Omega  \right\rangle $, where $\omega _{mn}  =  - \omega _{nm}$. In analogy to the Schmidt decomposition, it can be
proved that every $\left| \Psi \right\rangle $ can be represented in a
appropriately chosen basis in $C^N$ in a form of Slater decomposition \cite{schliemann},
\begin{eqnarray}
\left| \Psi \right\rangle =\frac 1{\sqrt{\sum_{i=1}^K\left| z_i\right| ^2}}%
\text{ }\sum_{i=1}^Kz_if_{a_1(i)}^{+}f_{a_2(i)}^{+}\left| \Omega
\right\rangle ,
\end{eqnarray}
where $f_{a_1(i)}^{+}\left| \Omega \right\rangle ,f_{a_2(i)}^{+}\left|
\Omega \right\rangle $, $i=1,\cdot \cdot \cdot ,K,$ form an orthonormal
basis in $C^N$. The number of the nonvanishing coefficients $z_i$ is called
the Slater rank, which is then used for the entanglement measure. With
similar technique, the case of two-boson system is studied by Li \cite{li} 
and Pa$\check{s}$kauskas
\cite{paskauskas}.

Gittings \cite{gittings} put forward three desirable properties of any
 entanglement measure:
(a) Invariance under local unitary transformations; (b) Non invariance under 
non-local unitary transformations; (c) 
Correct behavior as distingusishability of the subsystems is lost. 
These requirements make the distinction between 
one-particle unitary transformation and one-site unitary transformations 
become relevant. As claimed in 
Ref. \cite{gittings}, a natural way achieving this distinction is to use a 
basis based upon sites rather on particles. 
Through Gittings' investigation, it is shown that all the above-discussed 
entanglement measures fail the tests of 
the three criteria. Only the Zanardi's measure \cite{zanardi} survives, 
which is given in Fock space as the von 
Neuman entropy, namely,
\begin{equation}
 E_j =  - Tr\rho _j \ln \rho _j, \rho _j  = Tr_j \left| \psi  \right\rangle 
 \left\langle \psi  \right|, \\ 
\label{entanglement} 
\end{equation}
where $Tr_j$ denotes the trace over all but the $jth$ site and $\psi$ is 
the antisymmetric wave function of the 
studied system. Hence $E_j$ actually describes the entanglement of the 
$jth$ site with the remaining sites. This 
measure is well operational numerically and will be used in our 
lattice systems, which uses the basis based upon 
sites from the start. A generalization of this one-site 
entanglement is to define an entanglement between one L-site 
block with the rest of the systems \cite{vidal},
\begin{equation}
E_L  =  - Tr(\rho _L \log _2 \rho _L ).
\end{equation}
In this paper, this idea will be pursued together 
with the real-space renormalization group (RG) method.

\section{Entanglement and  quantum phase transition}

Quantum phase transition (QPT) is a qualitative change of some 
physical properties in a quantum many-body system 
as some parameter in the Hamiltonian is 
varied \cite{sachdev}. At the critical point when QPT happens, 
a long-range correlation can develop in the system. 
QPT is caused by quantum fluctuations at the absolute zero of temperature and 
is a pure quantum effect. Similarly entanglement is 
also a unique quantum property with non-local states of two or
more quantum systems superposed with each other. 
Hence it is inviting  to study the  relationship between them. 
Recently it has been argued that the property 
responsible for the long-range correlation in QPT is entanglement 
\cite{osborne}. So far, there have been some 
efforts  in this direction, such as the analysis 
of the  XY model about the single-spin entropies and two-spin 
quantum correlations \cite{osborne1}, the entanglement 
between a block of $L$ contiguous sites and the rest of the 
chain \cite{vidal} and also the scaling of entanglement
 near QPT \cite{osterloh}. But because there is still no 
analytical proof to validate the above speculation, 
the role played by the entanglement in quantum critical 
phenomena remains elusive. Generally speaking, 
there are existing  at least two difficulties in resolving this issue. 
First, until now, only two-particle entanglement 
is well explored. How to quantify the multi-particle entanglement 
is not clear.  Second, QPT closely relates to 
the notorious many-body problems, which is almost intractable 
analytically except in some special toy models, 
such as the Ising model and the XY model. Until now,
the only effective and accurate way to deal with 
QPT in critical region is the density-matrix RG method 
\cite{white}. Unfortunately, it is only efficient 
for one-dimensional cases because of the much more complicated 
boundary conditions for two-dimensional situation. 

In order to investigate the entanglement behavior in QPT for 
two-dimensional cases, we  used the original 
real-space RG technique \cite{realspace}. This method 
contains uncontrollable approximations. But by using bigger 
block size while at the same time comparing some of the obtained 
results with those already known from other methods 
to check the errors, we can still get reliable conclusions. 
For  the entanglement measure, we  used the Zanardie's measure. 
It is in essence a bipartite measure, but 
since most of the present quantum computation applications are 
concerning two-particle entanglement, it is still 
very meaningful and informative to carry out the study with it.

\section{Calculation Method}

We have shown previously that for a half-filled 
triangular quantum lattice, there exists a 
metal-insulator phase transition with the ratio of 
electron repulsion and hopping term to be the tuning parameter 
\cite{wang1, wang2, wang3}. In the following, we
 will use the same model and technique to explore the scaling
properties of the entanglements. 

The model we use is the Hubbard model with the Hamiltonian,

\begin{eqnarray}
H&=&-t\sum_{<i,j>,\sigma }[c_{i\sigma }^{+}c_{j\sigma }+H.c.]\nonumber \\
&&+U\sum_{i}(\frac{1}{2}-n_{i\uparrow 
})(\frac{1}{2}-n_{i\downarrow })
+K\sum_{i}I_{i} 
\end{eqnarray}
where $t$ is the nearest-neighbor hopping term, $U$ is the local repulsive
interaction and $K=-U/4$ and $I_i$ is the unit operator. $c_{i\sigma}^{+}(c_{i\sigma })$ creates(annihilates) 
an electron with spin $\sigma $ in
a Wannier orbital located at site $i$; the corresponding number operator is $%
n_{i\sigma }=c_{i\sigma }^{+}c_{i\sigma }$ and $<$ $>$ denotes the
nearest-neighbor pairs. H.c. denotes the Hermitian conjugate.
The order parameter for metal-insulator transition(MIT) is the charge gap 
defined by $\triangle 
_{g}=E(N_{e}-1)+E(N_{e}+1)-2E(N_{e})$, where $E(N_{e})$ denotes the lowest 
energy for a $N_{e}-$electron system. In 
our case, $N_{e}$ is equal to the site number $N_{s}$ of the lattice. 
We already know that at the critical point 
$(U/t)_c=12.5$, there will be a quantum phase transition, i.e. MIT,
 signature by a jump of $\Delta _g$ from Zero to 
non-zero. Unlike the charge gap calculated from the energy levels, 
the Zanardi measure of the entanglement is 
defined upon the wave function corresponding to $E(N_e)$ instead,
 as show in Eq. (\ref{entanglement}). Before 
showing how to get the entanglement from RG method, let us first 
briefly review the multi-staged real-space RG 
method.

The essence of real-space RG method is to map the original Hamiltonian to a
new Hamiltonian with much fewer freedoms which keeps the physical quantities
we are interested in unchanged \cite{realspace}. The map can be iterated until the final
Hamiltonian can be easily handled. The crucial step in this method is how
to related the parameters between the old and the new Hamiltonians. This can
be realized by dividing the original lattice into blocks and then build the
new Hamiltonian upon blocks, namely regard each block to be an effective
site.

Fig.1 shows schematically the hexagonal block structure we used in our
calculations and the coupling between blocks. For each block, we solve it
numerically in the subspaces of 6 electrons with 3 spin up and 3 spin
down, 7 electrons with 4 spin up and 3 spin down, 7 electrons with 3 spin up
and 4 spin down, and 8 electrons with 4 spin up and 4 spin down. In all
the subspaces, we keep the lowest-energy non-degenerate state, which is also required to
belong to the same irreducible representations of $C_{6\nu }$ symmetry
group. It should be mentioned here that if the there is degeneracy then one
possible solution is  to average over the degenerated states. 
The kept states will then be taken as the 4 
states for an effective site, $\left| 0\right\rangle ^{^{\prime }}$, $\left| \uparrow \right\rangle
^{^{\prime }}$, $\left| \downarrow \right\rangle ^{^{\prime }}$, $\left|
\uparrow \downarrow \right\rangle ^{^{\prime }}$.
 If denoting the energies corresponding to the first two states by $
E_{1}, E_{2},$ after some intensive calculations, we can obtain the new
Hamiltonian on the effective lattice, which has the same structure as the
original Hamiltonian, i.e.,
\begin{eqnarray}
H^{\prime } &=&-t^{\prime }\sum_{<i,j>,\sigma }[c_{i\sigma }^{\prime
+}c_{j\sigma }^{\prime }+H.c.]  \nonumber \\
&&+U^{\prime }\sum_{i}(\frac{1}{2}-n_{i\uparrow }^{\prime })(\frac{1}{2}
-n_{i\downarrow }^{\prime })+K^{\prime }\sum_{i}I_{i},
\label{Eq.4}
\end{eqnarray}
where the prime $^{\prime }$ denotes the operator action upon the block
states and
\begin{eqnarray}
t^{\prime } &=&3 \lambda ^{2}t, \\
U^{\prime } &=&2(E_{1}-E_{3}), \\
K^{\prime } &=&(E_{1}+E_{3})/2.
\end{eqnarray}
with $\lambda=<-\sigma ,\sigma |_p^{\prime }c_{i^{(p)}\sigma }^{+}|_p-\sigma >_p^{\prime
}=<\sigma |_p^{\prime }c_{i^{(p)}\sigma }^{+}|0>_p^{\prime }$, where $\sigma$ 
denotes $\uparrow$ or $\downarrow$. 
The above equations are the so-called RG flow equations. Usually, they are
iterated until we get the fixed point. But besides the fixed points, we can get more if 
we keep track of the results from each RG iteration.
For example, if we stop the RG flow at the first iteration, then
obtained $t^{\prime}$ and $U^{\prime}$ will be parameters of an 
effective 7-site hexagonal block mapped from the 
original system of $7^{2} $ sites. Because the hexagonal 
block Hamiltonian can be solved numerically,
then the physical quantities we have interest for a system 
containing 49 sites can be obtained. In this way, we can 
study the system of the size $7^{2},7^{3},7^{4},7^{5},7^{6}.....$. 
This is the so-called multi-stage real-space RG 
method, which is well adapted to start the finite-size scaling analysis. 
 For entanglement, because at the $ith$ 
iteration, the effective site involves $7^i$ starting sites, 
what we are calculating by this procedure is equivalent 
to get the entanglement between the $L$-site block and the 
surrounding seven L-site blocks. As $L$ increases with the 
RG flow, the scaling of the entanglement with $L$ comes out readily.

\section{Results and discussions}

In Fig. 2 we show the results of the entanglement as a function 
of $(U/t)$ for different system sizes.
The crossing point in Fig.2(a) represents the critical value of
 $U/t$, which is shown to be $(U/t)_c=12.5$. It is 
exactly equal to the critical value for metal-insulator 
transition (MIT) when the same order parameter $U/t$ is used 
\cite{wang1,wang2,wang3}. In Fig.2(b), with proper scaling, 
all the curves in Fig.2(a) collapse onto one curve, which 
can be expressed as
\begin{eqnarray}
E_7 = f(qN^{\frac{1}{2}})
\end{eqnarray}
where $q=U/t-(U/t)_c$ measures the deviation distance of the 
system away from the critical state. By the 
one-parameter scaling theory, near the phase transition point, $E_7$ can written as
\begin{eqnarray}
E_7 =q^{y_E} f(\frac{L}{\xi })
\end{eqnarray}
where $\xi  = q^{ - \nu }$ is the correlation length of the system
w here the critical exponent
 $\nu$. 
Hence,
\begin{eqnarray}
E_7 = q^{y_E } f(Lq^\nu ) = q^{y_E } f(N^{\frac{1}{{2\nu}}} q),
\end{eqnarray} 
where we used  $N=L^2$ for the two-dimensional systems. Now we can have,
\begin{eqnarray}
y_E=0, \nu=1
\end{eqnarray}
It is interesting to note that here we have acquired the same $\nu$ 
as in the study of MIT. This shows the great 
consistency of the results since the critical exponent $\nu$ is only 
dependent upon the inherent symmetry and 
dimension of the investigated system.

Another significance of the results lies in the finding that in the 
metal state, the system is highly entangled with 
$E_7=2$ while in the insulating state, the system is partly entangled 
with $E_7=1$. Because the reduced density 
matrix $\rho _7$ is four dimensional, the maximally entangled state 
can be written as
\begin{eqnarray}
\rho _7=\left( 
\begin{array}{llll}
\frac 14 &  &  &  \\ 
& \frac 14 &  &  \\ 
&  & \frac 14 &  \\ 
&  &  & \frac 14
\end{array}
\right),
\end{eqnarray}
in the basis $\left| 0 \right\rangle ,{\rm  }\left|  \uparrow  \right\rangle ,
{\rm  }\left|  \downarrow  
\right\rangle ,{\rm  }\left| { \uparrow  \downarrow } \right\rangle $. 
The related entanglement is $E_7 =  - 
\sum\limits_{i = 1}^4 {\frac{1}{4}\log _2 \frac{1}{4}}  = 2$. 
Unlike the metal state, in which the sites can have 
equal probability in any of the four basis states, 
the insulating states should be expected to have electrons 
showing less mobility. From the calculations, we 
know that in the insulating state, 
\begin{eqnarray}
\rho _7=\left( 
\begin{array}{llll}
0 &  &  &  \\ 
& \frac 12 &  &  \\ 
&  & \frac 12 &  \\ 
&  &  & 0
\end{array}
\right), 
\end{eqnarray}
which means that the central site has equal probability to be in $\left|  
\uparrow  \right\rangle $, $\left|  
\downarrow  \right\rangle $ and no occupation in $\left|  0  \right\rangle $, $\left| 
 \uparrow \downarrow  
\right\rangle $. This well coincides with the above expectations. The corresponding 
entanglement is then $E_7 =  - 
\sum\limits_{i = 2}^3 {\frac{1}{2}\log _2 \frac{1}{2}}  = 1$.

All the above discussions are confined to the entanglement between 
the central site and its surrounding sites. 
Because the central site is a very special site showing the highest 
symmetry in the block, one may wonder what can 
happen to the neighbor sites, for example, the entanglement between 
site 1 and the rest 6 sites. To answer this 
question, the same calculations are conducted and the results are 
presented in Fig.3. 
Following the same procedures as for site 7, we can get the 
transition point and the critical exponents. It is no 
surprise that the same values are obtained. The only difference
 is that in the metal state, the maximal entanglement 
is a little less than 2 and the minimal one is a little less than 1.
 This can be explained by the asymmetric 
position of site 1 in the block. 

It should be mentioned that the calculated entanglement here has a corresponding 
critical exponent $y_E=0$. This 
means that the entanglement is constant at the critical point over all sizes 
of the system, which can be well seen 
from Fig.2(a) and Fig.3(a) also. But it is not a constant over all values of $U/t$.
 There is an abrupt jump across 
the critical point as $L \to \infty $, which can be easily seen in Fig.2(b) 
and Fig.(3)b. If we divide the regime of 
the order parameter into non-critical regime and critical regime, 
the results can be summarized as follows. 1) In 
the non-critical regime, i.e. $U/t$ is away from $(U/t)_c$, as $L$ increases, 
the entanglement will saturate finally 
onto two different values depending upon the sign of $U/t-(U/t)_c$. 2)
At the critical point, the entanglement is 
actually a constant independent of the size $L$. These properties are 
qualitatively different from the single-site 
entanglement discussed by Osborne \cite{osborne1}, where the entanglement 
with Zanardi's measure increases from zero 
to the maximum at the critical point and then decreases again to zero as 
the order parameter $\gamma$ for $XY$ mode 
is tuned. The peculiar properties of the entanglement we have found here 
can be of potential interest to make an 
effective ideal entanglement "switch". For example, with seven blocks of 
quantum dots on triangular lattice, the 
entanglement among the blocks can be regulated as $"0"$ or $"1"$ almost 
immediately once the tuning parameter 
$\frac{U}{t}$ crosses the critical point $12.5$. The switch errors will 
depend upon the size of the blocks. Since it 
has already been a well-developed technique to change $\frac{U}{t}$ for 
quantum dot lattice, the above scheme should 
be well workable. To remove the special confinement we have made upon 
the calculated entanglement, namely only the 
entanglement of site 1 and site 7 with the rest sites are considered,  
in the following, we will prove that what we 
are switching here is actually the average pairwise entanglement between 
the seven blocks.

Let $E_{ij}$ denotes the entanglement between $ith$ and $jth$ site. Then the 
total two-site entanglement $E_{tot}$ 
can be obtained, i.e.  
\begin{eqnarray}
E_{tot}  = \sum\limits_{i = 2}^7 {E_{1i} }  + \sum\limits_{i = 3}^7 {E_{2i} }  + 
\sum\limits_{i = 4}^7 {E_{3i} } 
\nonumber \\  
+ \sum\limits_{i = 5}^7 {E_{4i} }  + \sum\limits_{i = 6}^7 {E_{5i} }  + E_{67}.
\label{eq1} 
\end{eqnarray}
From the symmetry of the sites, we know that
\begin{eqnarray}
\begin{array}{l}
 E_{ij}  = E_{ji} , \\ 
 E_{13}  = E_{24}  = E_{35}  = E_{46} ,{\rm  }E_{14}  = E_{25}  = E_{36} , \\ 
 E_{17}  = E_{27}  = E_{37}  = E_{47}  = E_{57}  = E_{67} . \\ 
 \end{array}
\end{eqnarray}
By substituting the above equalities into Eq. (\ref{eq1}), we have,
\begin{eqnarray}
E_{tot}  = 6E_{17}  + 3(2E_{12}  + 2E_{13}  + E_{14} ).
\end{eqnarray}
Because 
\begin{eqnarray}
E_7  = 7E_{17} ,{\rm   }E_1  = 2E_{12}  + 2E_{13}  + E_4 
\end{eqnarray}
We finally have $E_{tot}  = 6E_7  + 3E_1 $. The average 2-site entanglement
 is $E_{average}  = \frac{{E_{tot} }}{7} 
= \frac{1}{7}(2E_7  + E_1 )$. Because
\begin{eqnarray}
\begin{array}{l}
 E_7  = f(qN^{\frac{1}{2}} ), \\ 
 E_1  = g(qN^{\frac{1}{2}} ), \\ 
 \end{array}
\end{eqnarray}
then we should have $E_{average}  = h(qN^{0.5} )$. This tells us that 
the average pairwise entanglement also has the 
properties shown in Fig.2 and Fig.3. 

Until now, almost all the finite-scaling work about the entanglements are 
done in one-dimensional cases, such as 
Ising model with transverse magnetic field or XY model. And different scaling 
properties of the entanglement are 
found there. In \cite{lak}, the one-dimensional Harper Hamiltonian is investigated.
 The average concurrence shows 
different scaling properties in the metal and insulating states. In \cite{vidal},
 Vidal calculated the entanglement 
with Zanardi measure between a $L$-site block with all the rest spin chains using
 XY model. And at the critical
point, the entanglement demonstrates a logarithm dependence over the size $L$. 
Similar results are also found by 
Osterloh \cite{osterloh} in the concurrence derivative over the order parameter 
$\lambda$. None of the above work 
has reported the switching properties of the entanglement scaling over size. 
Hence it seems that the dimension of 
the investigated systems seem to play an important role also in the entanglement 
scaling, just as expected from QPT 
viewpoint.

\section{Summary}

In summary, by using a multi-stage real-space renormalization group method, 
the scaling properties of the
entanglement with the Zanardi's measure over the block size when QPT happens
 has been investigated in details. The 
critical exponent $\nu=1$ has been found, which coincides well with our 
previous work in studying a quite different 
physical property, the charge gap. When the block size $L \to \infty$, 
the entanglement shows an abrupt change when 
the tuning parameter crosses the phase transition point. This property 
might be well applied to make an "entanglement 
switch".

\begin{acknowledgments}

We would like to acknowledge the financial support of the Office 
of Naval Research(N00014-97-2-0192)

\end{acknowledgments}

\newpage
\begin{figure}[tbp]
\includegraphics{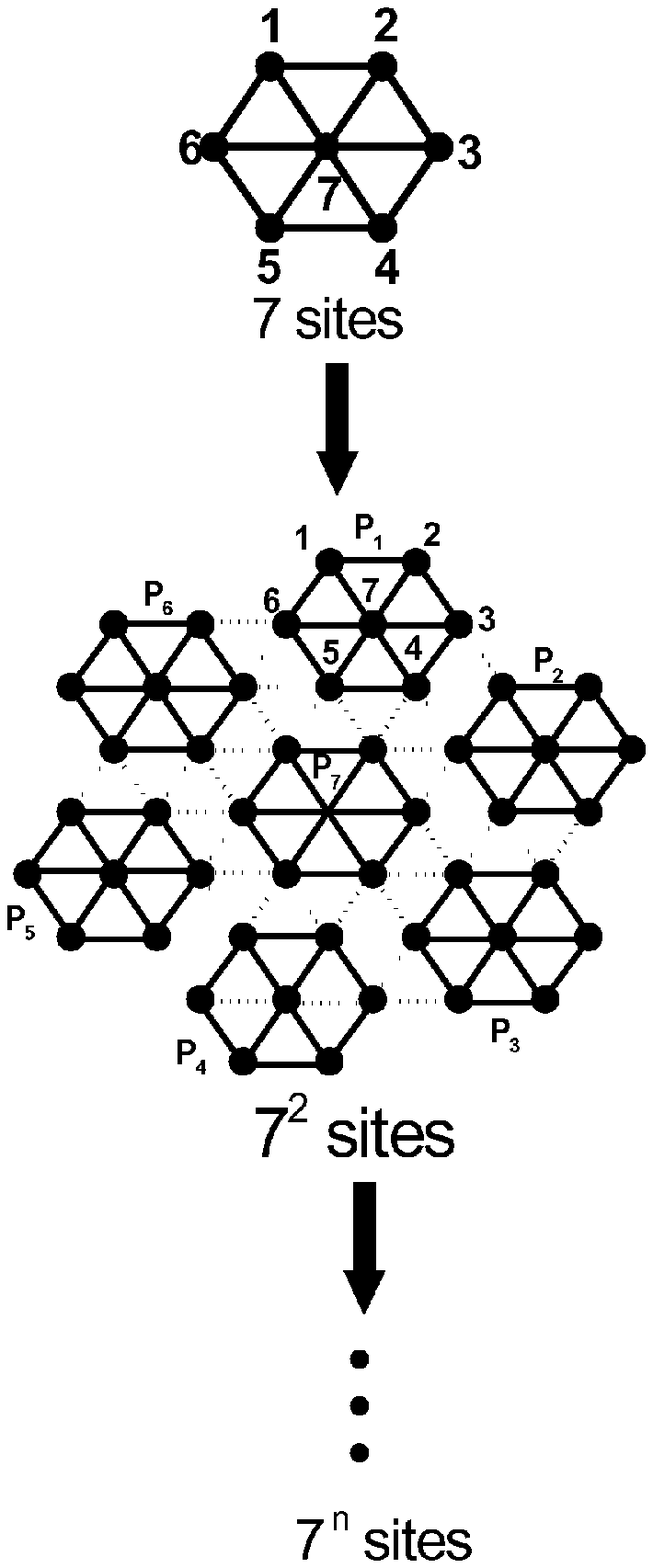}
\caption{ Schematic diagram of the triangular lattice with hexagonal blocks.
The hexagonal block is used for the renormalization process. The iteration direction
is shown by the arrow. The dotted lines represent the interblock interactions and solid line intrablock ones.}
\end{figure}
\newpage

\begin{figure}[tbp]
\includegraphics{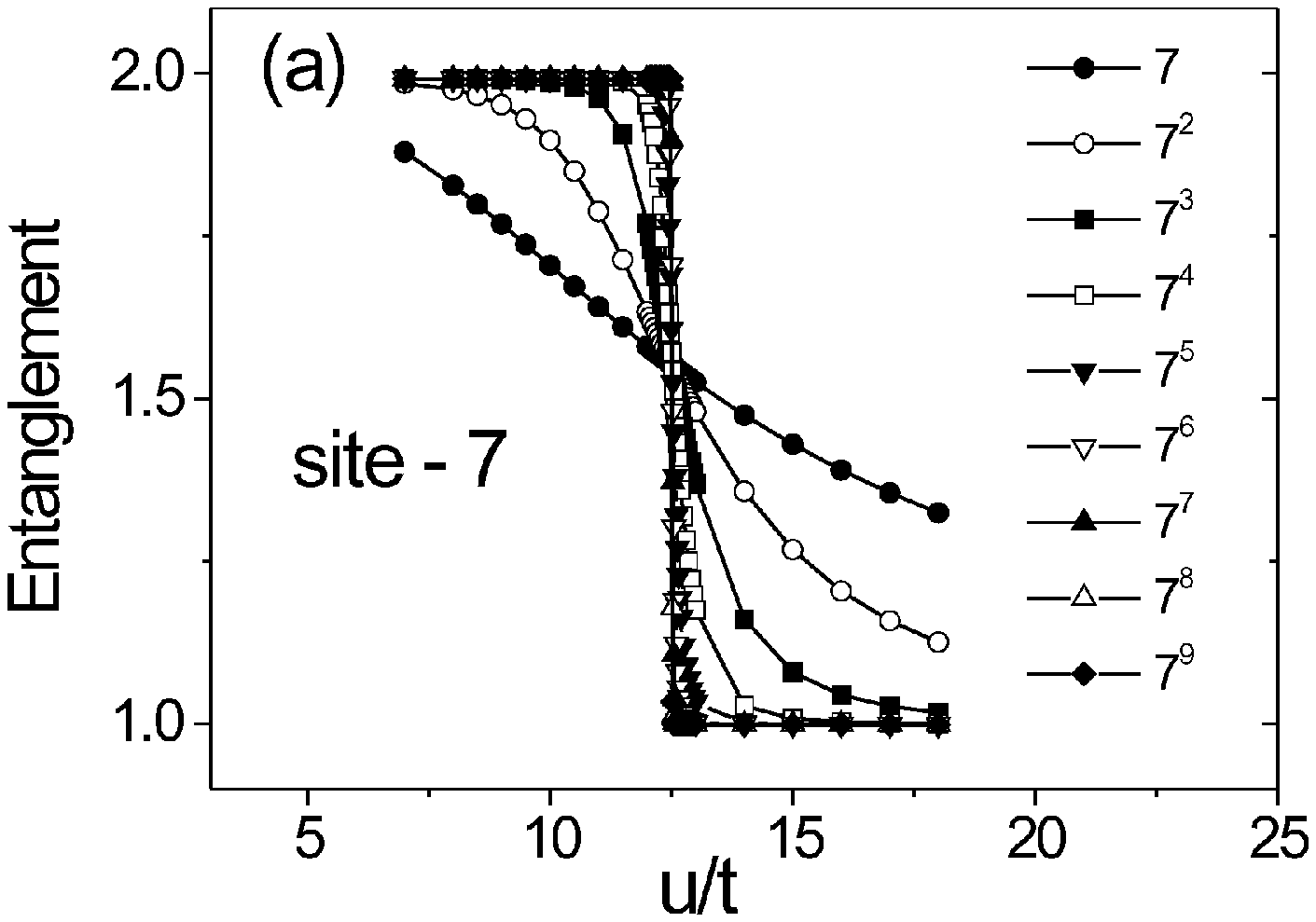}
\caption{Variations of the entanglement between site 7 and the neighboring sites (See Fig.1) against the ratio of on-site electron interaction $U$ to the hopping term $t$ for different
system size, i.e. the number of sites: $7$(solid circle), $7^2$(open circle), $7^3$(solid square), $7^4$(open square) $7^5$(solid down triangle), $7^6$(open down triangle), $7^7$(solid up triangle), $7^8$(open up triangle), $7^9$(solid diamond)}
\end{figure}

\newpage
\begin{figure}[tbp]
\includegraphics{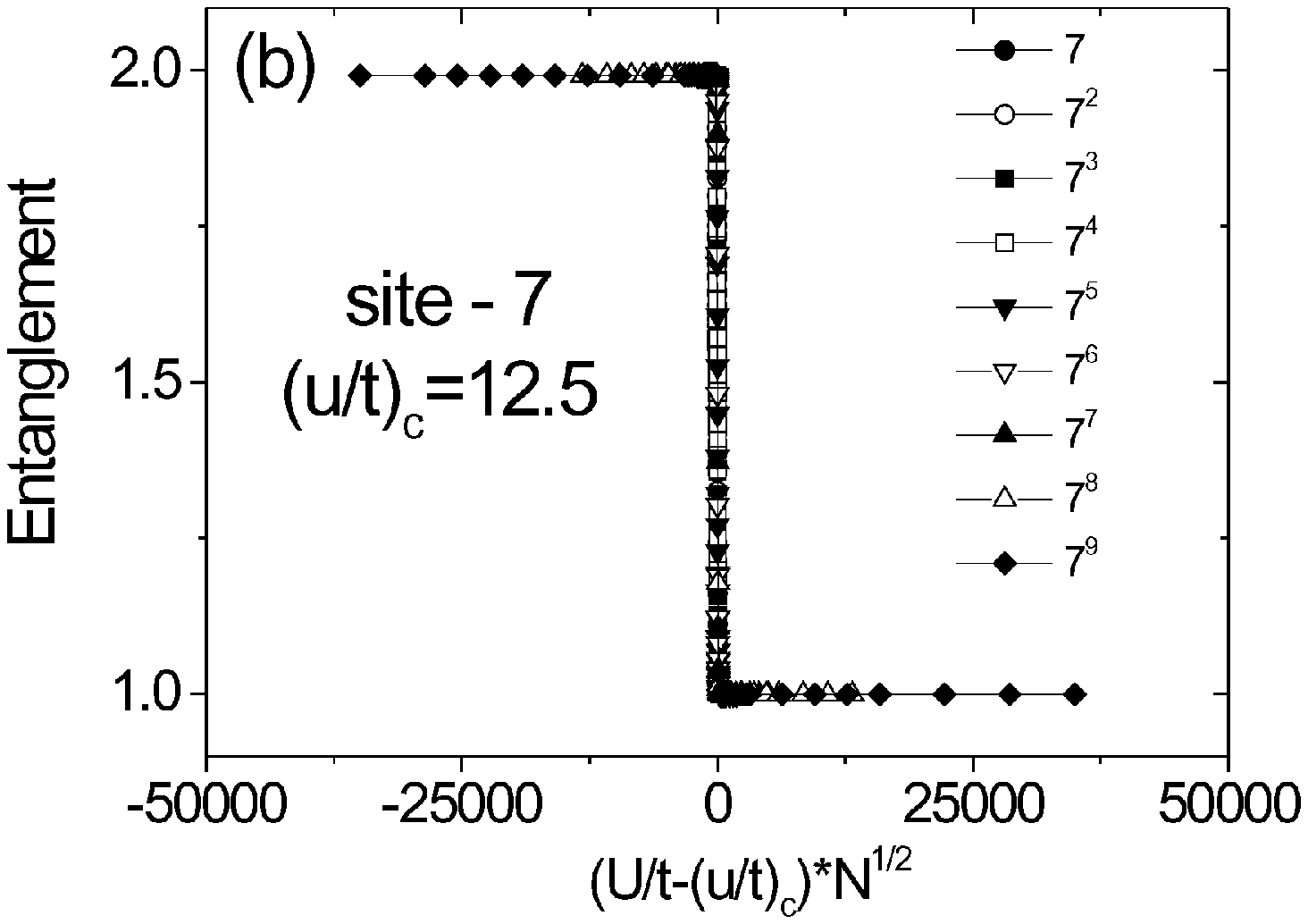}
\caption{Variations of the entanglement between site 7 and the neighboring sites (See Fig.1) against the ratio of on-site electron interaction $U$ to the hopping term $t$ for different
system size, i.e. the number of sites: $7$(solid circle), $7^2$(open circle), $7^3$(solid square), $7^4$(open square) $7^5$(solid down triangle), $7^6$(open down triangle), $7^7$(solid up triangle), $7^8$(open up triangle), $7^9$(solid diamond)}
\end{figure}

\newpage
\begin{figure}[tbp]
\includegraphics{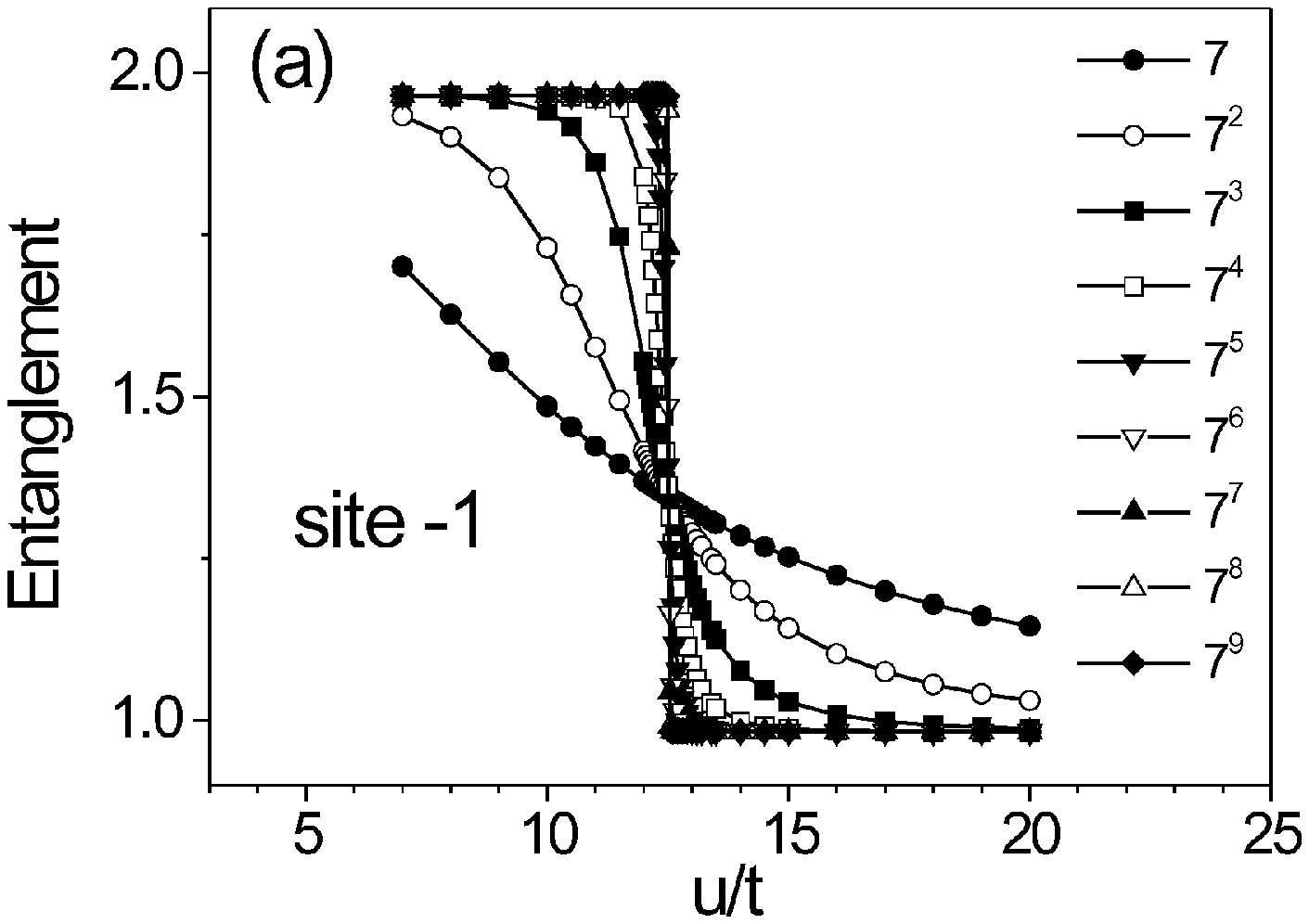}
\caption{The same as in Fig.2, but for the entanglement between site 1 and the rest 6 sites of the block.}
\end{figure}
\newpage
\begin{figure}[tbp]
\includegraphics{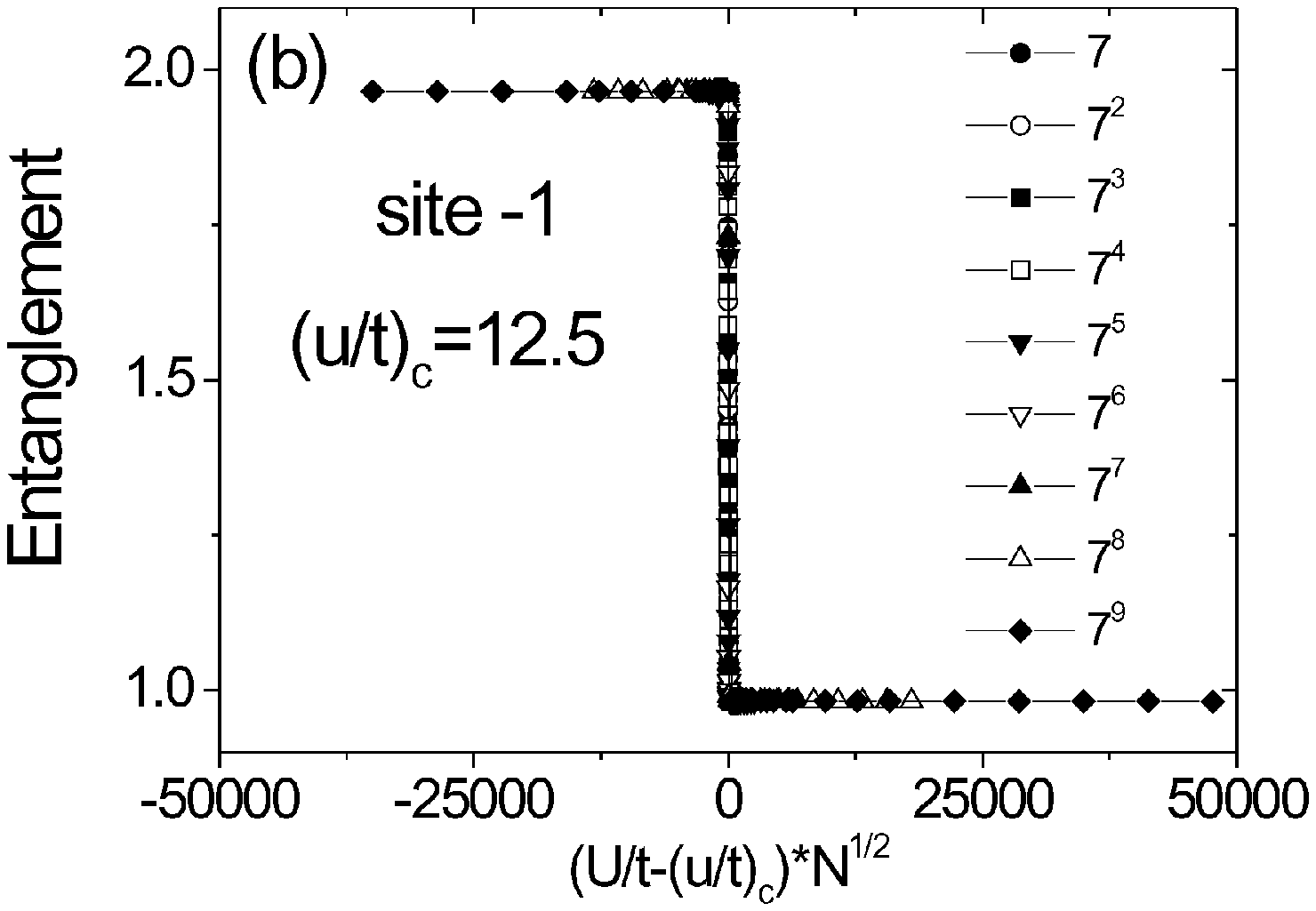}
\caption{The same as in Fig.2, but for the entanglement between site 1 and the rest 6 sites of the block.}
\end{figure}

\end{document}